
\input harvmac
\noblackbox
%
%

\def\NP{{\it Nucl. Phys.\ }}

\def\PL{{\it Phys. Lett.\ }}
\def\PR{{\it Phys. Rev.\ }}
\def\PRL{{\it Phys. Rev. Lett.\ }}
\def\CMP{{\it Comm. Math. Phys.\ }}

\font\tiau=cmcsc10
\baselineskip 12pt
\Title{\vbox{\baselineskip12pt \hbox{hep-th/9410215}
\hbox{UCSBTH-94-42} }}
{\vbox{\hbox{\centerline{\bf STRINGS NEAR A RINDLER  OR
BLACK HOLE HORIZON
}}}}
\centerline{\tiau David A. Lowe\foot{lowe@tpau.physics.ucsb.edu}}
\centerline{\tiau and}
\centerline{\tiau Andrew Strominger\foot{andy@denali.physics.ucsb.edu}}
\vskip.1in
\centerline{\it Department of Physics}
\centerline{\it University of California}
\centerline{\it Santa Barbara, CA 93106-9530}
\vskip .5cm
\noindent
Orbifold techniques are used to study bosonic, type II and
heterotic strings in Rindler
space at
integer multiples $N$ of the Rindler temperature, and near a black hole horizon
at integer multiples of the Hawking temperature, extending earlier results of
Dabholkar. It is argued that a Hagedorn transition occurs nears the horizon
for all $N>1$.

\Date{October, 1994}
\lref\variat{M. Ba\~nados, C. Teitelboim and J. Zanelli, \PRL {\bf 72} (1994)
957; S. Carlip and C. Teitelboim, preprint IASSNS-HEP-93-84,
gr-qc/9312002, C. Callan and F. Wilczek, \PL {\bf B333} (1994) 55.}
\lref\atick{J.J. Atick and E. Witten, \NP {\bf B310} (1988) 291;
W. Fischler and J. Polchinski unpublished.}
\lref\dabhol{A. Dabholkar, Harvard preprint HUTP-94-A019,
hep-th/9408098; Caltech preprint CALT-68-1953, hep-th/9409158.}
\lref\callan{J.A. Bagger, C.G. Callan and J.A. Harvey, Nucl. Phys.
{\bf B278} (1986) 550.}
\lref\lat{I. Klebanov and L. Susskind, \NP {\bf B309} (1988) 175.}
\lref\haged{R. Hagedorn, {\it Nuovo Cim. Suppl. } {\bf 3} (1965) 147.}
\lref\dual{V. P. Nair, A. Shapere, A. Strominger and F. Wilczek,
\NP {\bf B287} (1987) 402.}
\lref\scatt{D.J. Gross, \PRL {\bf 60} (1988) 1229.}
\lref\susskind{L. Susskind and J. Uglum, \PR {\bf D50} (1994) 2700.}
\lref\lsuss{L. Susskind, {\it Some Speculations about Black Hole
Entropy in String Theory}, Rutgers University preprint RU-93-44,
hep-th/9309145.}
\baselineskip 12pt

\newsec{Introduction}

The usual perturbative
description of string theory as a theory of quantum strings
propagating in a fixed spacetime background fails miserably
at the Planck scale, just when string theory is supposed to be telling us
something really interesting. It appears that the proper description
must be fundamentally different and does not involve either strings or
spacetime. Evidence for this, arising from calculations involving
strings in various kinds of extreme circumstances, is given by
the existence of the
Hagedorn transition \haged, $R\rightarrow 1/R$ duality in spacetime \dual,
lattice formulations \lat\ and symmetries in
high energy scattering \scatt. These few concrete calculations have
served as precious guideposts in our attempts to understand planckian
string theory.

It is important to develop additional guideposts. One
such potential guidepost is the behavior of strings near a
black hole horizon. This may also bear on the information
problem \refs{\lsuss,\susskind}.
In the limit that the black hole becomes very large, its horizon
approaches that of the Rindler wedge, and one is led to the simpler
problem of strings near the Rindler horizon.

A complete understanding of strings in Rindler space has been elusive.
This subject has previously been considered in
\ref\sanchez{H.J de Vega and N. Sanchez, \NP {\bf B299} (1988) 818;
\PR {\bf D42} (1990) 3969; N. Sanchez, \PL {\bf B195} (1987) 160.}.
Ordinary quantum field theory in Rindler space is defined by constructing
annihilation and creation operators which are positive and negative frequency
with respect to Rindler time. Rindler time is proportional to the proper
time along a family of uniformly accelerating trajectories which fill the
Rindler wedge, with the proportionality constant equal to the
acceleration.
Of course the thermal Rindler state at Rindler temperature $T=1/2\pi$ is
equivalent to the Minkowski vacuum state at $T=0$ (restricted to the
Rindler wedge), so many of its properties can be trivially deduced,
even for the case of string theory\foot{Although
questions involving uniformly accelerated observers in this state
remain puzzling.}.
For example the energy density in this state vanishes everywhere. The
uniformly accelerated
Rindler observer explains this as due to a miraculous cancellation between the
Casimir energy of the Rindler vacuum and the thermal energy of the
gas of excited strings. These both increase in equal but opposite manners
near the Rindler horizon, the former due to the proximity of the boundary,
the latter due to the blue-shifting of the local temperature.

It is much more non-trivial to understand Rindler strings at
$T\neq 1/2 \pi$ (including $T=0$, the Rindler vacuum).
In this case the cancellation will not occur,
because the thermal energy will change with the temperature but the
vacuum energy will not. Since the local temperature diverges as the
inverse distance to the horizon, one may expect a Hagedorn transition
near the horizon \ref\barbon{J.L.F. Barbon, PUPT-94-1478, hep-th/9406209}.
However, this
increasing temperature does not in itself necessarily imply a Hagedorn
transition because the size of the region in which the local
temperature exceeds the Hagedorn temperature is itself of
order the string scale. Thus a more careful analysis is required.

The general problem of Rindler strings at
$T\neq 1/2 \pi$ has not been solved. Indeed it is not
clear that there is {\it any} consistent description.
However for the special cases
$T= N/2\pi$, where $N$ is an integer greater than zero, the
partition function can be computed as a $Z_N$ orbifold. This orbifold
corresponds to a cone with deficit angle $\epsilon=2\pi (N-1)/N$.  We shall see
that tachyons appear (even for the superstring) with masses
$M^2=-2\epsilon/\pi$.
This indeed signals the
existence of a super-Hagedorn region near the horizon.

This same partition function also approximates the near-horizon behavior
of a black hole of arbitrary mass $M$ immersed in a heat bath
of temperature $N$ times the Hawking temperature $T_H=1/8 \pi M$.

The appearance of tachyons and a super-Hagedorn region
of course implies a breakdown of string perturbation
theory. What we have learned is that ill-understood
planckian string physics -- perhaps a genus-zero condensate\atick
-- is needed to describe physics near the horizon.

It has been suggested \refs{\lsuss,\variat} that the
entanglement entropy of
the quantum state outside a black hole horizon can be computed in string theory
as the derivative of the partition function with respect to the
deficit angle $\epsilon$ evaluated at $\epsilon = 0$.
It might seem that this suggestion would be foiled by the
genus-zero condensate which occurs for non-zero $\epsilon$.
In fact we shall argue -- using an analytic continuation of our
results at discrete values of  $\epsilon$ -- that this is not the case:
the uncalculable
genus-zero corrections to the partition function come in at order
$\epsilon^2$ and thus do not affect the entropy. This is perhaps
to be expected since other perturbative methods of computing the
entropy would not require consideration of strings on singular spaces.

In section 2 we compute the bosonic string partition function for
Rindler strings at $T= N/2\pi$.
Section 3 contains the type II superstring in the RNS formalism
and a discussion of the genus-zero condensate and its effect
on the calculation of the horizon entropy.
Section 4 describes the heterotic string.

Many of the results in sections 2 and 3 are contained in
a previous paper of Dabholkar \dabhol. This paper may be viewed as an
extension of his work with some differences in the
discussion and perspective.

\newsec{The Bosonic String}

We are interested in describing strings propagating on Rindler space
at finite temperature.
The thermodynamic partition function is obtained by performing
the path integral on the Euclidean continuation of this space.
The metric on this space may be written
in the following form
\eqn\rindmet{
ds^2 = r^2 d\theta^2 + dr^2 + d \vec x^2~,
}
where the imaginary  time coordinate $\theta$ is periodically
identified, with period $\beta$, the inverse temperature.

This metric is the analytic continuation of a cosmic string metric
\eqn\cosmicmet{
ds^2 = -dt^2 + r^2 d\theta^2 + dr^2 + d\vec{ \tilde x}^2~,
}
in which the $x_2$ coordinate is continued to $t=-ix_2$. The mass
of the cosmic string introduces a conical singularity at $r=0$ with
 non-trivial deficit angle $2\pi-\beta$. Quantum field theory
on cones has been studied in
\ref\dowker{J.S. Dowker, p 251, {\it The formation and evolution of
cosmic strings}, ed. G. Gibbons, S.W. Hawking and T. Vachaspati,
Cambridge University Press, 1990, and references therein.}.
Strings propagating on spaces with cosmic string type singularities
have been previously considered in \refs{\dabhol, \callan}.
The cosmic string continuation of \rindmet\
allows us to fix light-cone gauge, which will be helpful
when we consider the spectrum of the string theory.

In the following we will be considering critical string theories
obtained by orbifolding flat space
\ref\dixon{L. Dixon, J. Harvey, C. Vafa and E. Witten, \NP {\bf B261}
(1985) 678; \NP {\bf B274} (1986) 285.}. These models satisfy on-shell
conformal invariance ( unlike the off-shell prescription proposed in
\susskind\ to continue the
the partition function to non-trivial $\beta$ ). A consequence
of this will be that the genus-one partition functions we obtain
may be written as integrals of modular invariant expressions over
the fundamental domain. Because the UV region is excluded from
these integrals, all our expressions will be UV finite.
Unfortunately, the only known modular invariant theories are
$Z_N$ orbifolds of flat space, which restricts the deficit
angle to a discrete set of values $2\pi(1-1/N)$.

Consider the critical closed bosonic string on the Euclidean orbifold
$R^{24} \times R^2/ Z_N$. It is convenient to represent $X_1$ and $X_2$ as a
complex coordinate, $X= (X_1 + i X_2)/ \sqrt{2}$ and
$\bar X = (X_1 - i X_2)/ \sqrt{2}$.
The Hilbert space of the string states splits up into an untwisted sector
and a number of twisted sectors labeled by the integer $k=0,\cdots,N-1$.
The boundary condition
in the $k$'th twisted sector is then
\eqn\twbc{
X(\sigma + 1) = e^{2\pi i k/N} X(\sigma).
}
This leads to the following mode expansion for the $X$ field
\eqn\modex{
X= \delta_{k,0}x +
{i \over 2} \sum_n { {\alpha_{n+k/N} } \over {n+ k/N}}
e^{2i\pi (n+k/N) \sigma } + {i \over 2} \sum_n
{ {\tilde \alpha_{n-k/N} } \over {n- k/N}}
e^{-2i\pi (n-k/N) \sigma }~,
}
in units where $\alpha'=1/2$. The oscillators satisfy the
commutation relations
\eqn\ocomm{
\eqalign{
[ \alpha_{m+k/N} , \bar \alpha_{n-k/N} ] &= (m+k/N) \delta_{m+n} \cr
[ \tilde \alpha_{m-k/N} , \bar {\tilde \alpha}_{n+k/N} ]
&= (m-k/N) \delta_{m+n} ~.\cr}
}
The twist in the boundary condition of the $X$ field introduces
a shift in the vacuum energy.
The contribution to the vacuum energy of $X$ and $\bar X$ is given by
$-1/2 (\eta^2 - \eta + 1/6)$, where $\eta = k/N$. The other
coordinates contribute a $-11/12$ to the vacuum energy, so the total
vacuum energy in the k'th twisted sector is $-1/2(\eta^2 - \eta +2)$.
Each twisted sector therefore contains one set of tachyon states with
$M^2 = 4(-\eta^2+\eta-2)$.

In the untwisted sector, states are labeled by the continuous
momenta $p$ and $\bar p$, together with momenta in the directions
transverse to the plane.
Note that there is a projection onto $Z_N$ invariant states,
so the allowed combinations of $p$ and $\bar p$ are reduced by a factor $N$.
In the twisted sectors, the $X$ field has no zero-mode, so states
are built on the vacuum state which has zero momentum in the $X_1$ and
$X_2$ directions. This means that
all states in the twisted sectors may be interpreted as states
localized near the tip of the cone, which may only propagate
in the transverse directions.

The orbifold partition function involves a sum over all the
twisted sectors, together with a projection onto $Z_N$ invariant states.
This is achieved by summing over the $N^2$ contributions ${\cal Z}_{k,l}$
where ${\cal Z}_{k,l}$ is the partition function for a single complex
boson on a torus with twisted boundary conditions
\eqn\twbc{
X(\sigma_1+1, \sigma_2) = e^{2\pi i k/N} X(\sigma_1, \sigma_2), \qquad
X(\sigma_1, \sigma_2+1) = e^{-2\pi i l/N} X(\sigma_1, \sigma_2)~.
}
Here $\sigma_1,\sigma_2$ are coordinates with period 1 which
parametrize the torus.
The partition functions ${\cal Z}_{k,l}$ are  then simply determinants
of Laplacians on the torus, subject to the above boundary
conditions.
\nref\alvarez{L. Alvarez-Gaume, G. Moore and C. Vafa, \CMP {\bf 106} (1986)
1.}%
These partition functions may be easily computed using the oscillators
defined in \modex,
\eqn\klpart{
{\cal Z}_{k,l} = \Biggl| { {\eta }\over {\theta\Bigl[ {{\ha+{k\over N}}
\atop {\ha+{l\over N}}} \Bigr] }} \Biggr|^2~,
}
for $k$ and $l$ not both zero.
Here $\theta\bigl[ {{\alpha} \atop {\beta}} \bigr]$ is the theta
function with characteristics $\alpha$ and $\beta$
\ref\mumford{D. Mumford, {\it Tata Lectures on Theta}, vol. I, Boston:
Birkh\"auser 1983.}.
The ${\cal Z}_{0,0}$ term involves
zero modes which yield a factor proportional to the
area of the plane $A_p$. One then obtains
\eqn\zzero{
{\cal Z}_{0,0} = {{A_p} \over { {\rm Im \tau}| \eta^2|^2}} ~.
}
This term is modular invariant on its own. The other ${\cal Z}_{k,l}$
mix under modular transformations.

The final partition function is then
\eqn\bpartit{
{\cal Z}_N = {1\over N}
{1\over { | \eta^2 {\bar \eta}^2 {\rm Im} \tau |^{11} }}
\sum_{k,l = 0}^{N-1} {\cal Z}_{k,l} ~,
}
and the genus one amplitude is
\eqn\toramp{
A_N = V_T \int_{\cal F} { {d^2 \tau} \over {({\rm Im} \tau)^2}}~
{\cal Z}_N~,
}
where ${\cal F}$ is the fundamental domain of the modular group
and $V_T$ is the 24-dimensional transverse volume.
The free energy of strings in Rindler space at finite temperature
is related to this via
\eqn\freef{
\beta F = - A_N~.
}

Since each twisted sector contains a tachyon, the free energy
\freef\ will diverge in the infrared.
The limit ${\rm Im}\tau \to \infty$ part of the
integral in the k'th twisted sector looks like
\eqn\asymdiv{
 -\int^{\infty}{{ d\tau_2}\over {\tau_2^2}} {{e^{-\pi M^2_k \tau_2/2}}\over
{\tau_2^{12} }}~,
}
where $M^2_k$ is the mass squared of the tachyon in the $k$'th twisted
sector. Thus the free energy in these models is negative infinity.

The presense of tachyons in the twisted sectors is  reminiscent
of the appearance of tachyonic winding states in string theory
in flat space at finite temperature as the Hagedorn temperature
is approached
\ref\kogan{Ya. I. Kogan, {\it JETP Lett.} {\bf 45} (1987) 709;
B. Sathiapalan, \PR {\bf D35} (1987) 3277.}. Indeed the twisted
sector states wind around the cone in the euclidean time direction.
They also lead to divergence of the thermodynamic partition
function and a breakdown of the canonical ensemble.
In
\atick\ it was argued that as the temperature approaches the Hagedorn
temperature, a first order phase transition should occur with
a large latent heat. At this phase transition a condensate of
the winding modes will form. This in turn leads to a
genus-zero contribution to the free energy.

Exactly the same kind of argument will apply here, with the
role of the winding modes replaced by the tachyonic fields
appearing in the twisted sectors. The infrared
divergence of \toramp\ is a consequence of expanding
around an unstable vacuum in the computation of the
free energy.
The correct procedure is to give the winding modes
expectation values which lead to stable solutions,
minimize the free energy.
In general one might also expect the dilaton and possibly other massless
fields to acquire expectation values. The problem
of finding this non-trivial minimum, and
describing the physics beyond this phase transition, appears to require
non-perturbative string physics and is at present intractable.

\newsec{Type II Superstring}

In this section, we consider the case of
type II superstrings to verify that a similar condensate forms in this case.
In the following, we use the covariant Ramond-Neveu-Schwarz formalism.
These theories have previously been considered in \dabhol\ using the
light--cone Green--Schwarz formalism.
In superconformal gauge, the worldsheet action for the RNS string is
\eqn\rnsact{
S= -{1\over {2\pi }} \int d^2\sigma~\bigl(
\del_\alpha X^\mu \del^\alpha X^\nu - i \bar \psi^\mu \rho^\alpha \del_\alpha
\psi^\nu \bigr) \eta_{\mu \nu}~,
}
where $\psi^\mu(\sigma)$ are Majorana--Weyl worldsheet spinors
transforming as a spacetime vector, i.e. in the same way as $X^\mu(\sigma)$.
As before, we combine $X_0$ and $X_1$ into a complex coordinate $X$,
and likewise combine $\psi^0$ and $\psi^1$ into a complex spinor $\psi$.
To construct physical states it is necessary to do a GSO projection,
which leads to a sum over spin structures on the worldsheet.

\subsec{Theories without spacetime fermions}

To begin with, we will ignore the level matching conditions which will
lead to theories with an infinite number of spacetime fermionic states.
In the $k$'th twisted sector ($k=0,\cdots, N-1$), the boundary conditions are
\eqn\fbcs{
\eqalign{
\psi(\sigma_1+1, \sigma_2) &= e^{2\pi i k/N} \psi(\sigma_1, \sigma_2)
\cr
\psi(\sigma_1+1, \sigma_2) &= -e^{2\pi i k/N} \psi(\sigma_1, \sigma_2)
\cr
X(\sigma_1+1, \sigma_2) &= e^{2\pi i k/N} X(\sigma_1, \sigma_2) \cr}
\qquad \eqalign{&{\rm (R)} \cr & {\rm (NS)} \cr & \cr}
}
where R refers to the Ramond sector, and NS refers to the Neveu--Schwarz
sector. The partition functions for the twisted fermions are easily
computed \alvarez, and lead to the following modular invariant
partition function for the orbifold
\eqn\fpartit{
{\cal Z}_N = {1\over {4N}}
{1\over {| \eta^2 {\bar \eta}^2 {\rm Im}\tau|^3 }}
\sum_{k,l=0}^{N-1} \sum_{\alpha,\beta}
 \Biggl| {{ \theta^3 \Bigl[ {\alpha \atop
\beta} \Bigr] \theta\Bigl[ {{\alpha+{k\over N}} \atop {\beta+{l \over N}}}
\Bigr]
} \over {\theta\Bigl[ {{ \ha+{k \over N}} \atop
{ \ha+{l \over N}}} \Bigr] \eta^3 }} \Biggr|^2 ~.
}
Here we are performing a diagonal sum over spin structures, labeled
by $\alpha,\beta=0,1/2$. This corresponds
to projecting onto states with $(-1)^F = +1$ in the (NS,NS) sector,
where $F$ is the sum of the fermion numbers of the left and right movers.
In the (R,R) sector the projection is also onto states with
$(-1)^F = +1$. The $k=l=0$ term should be interpreted as in \zzero.

In the usual type II string there is a separate GSO projection for
left and right movers which leads to spacetime fermions in the
(R,NS) and (NS,R) sectors. However, here these sectors are absent and
these models therefore contain only spacetime bosons.
The NS ground state is not projected out, so
tachyons are present both
in the untwisted and twisted sectors, with
$M^2 = 4(-1+k/N)$. Note that in the
Ramond sector, states could alternatively be projected onto
$(-1)^F=-1$ which would lead to the same partition function, but
a different spectrum. The theories with $N=1$ have been considered
previously by Seiberg and Witten
\ref\seiberg{N. Seiberg and E. Witten, \NP {\bf B276} (1986) 272.}.
Using the Riemann theta identity \mumford\ it may be shown that
\fpartit\ agrees
with \dabhol\ (equation (2.21) for $N$ even, and equation (2.22) for $N$ odd).

\subsec{Theories with spacetime fermions}

To obtain consistent theories with spacetime fermions some
care is required.
The action of the discrete group $Z_N$ on the worldsheet fields must
be chosen in such a way that there is level matching between
the right moving NS sector and the left moving R sector (and vice versa), to
ensure an infinite number of fermionic states appear in the spectrum.
In addition, the GSO projection must be chosen in a manner
consistent with modular invariance. It should be noted that any theory
with non-trivial deficit angle will necessarily break spacetime
supersymmetry.

Let us consider the contributions to the partition function
coming from the worldsheet fermion determinants in the (R,NS) sector.
Inserting the GSO projection, this corresponds to
${1\over 4}{\rm Tr}( 1+(-1)^{F_L})(1-(-1)^{F_R})
 q^{H_L} {\bar q}^{H_R}$ where the
left-moving states are Ramond fermions twisted by $h\in Z_N$,
and the right-moving states are Neveu-Schwarz fermions twisted by $h$.
$H_L$ ($H_R$) denotes the left-(right-)moving hamiltonian for the
fermions, and $F_L$ ($F_R$) denotes the left-(right-)moving worldsheet
fermion number. Because $h^N=1$, the boundary conditions on the fermions are
unchanged under a modular transformation $\tau \to \tau+N$, so the
trace should not pick up a phase.
This will be true, provided that for an arbitrary state which
is invariant under the GSO projections,
\eqn\levm{
N(E_L -E_R) = 0~ {\rm mod}~1.
}
We take the action of the group element $h$ on the fermions to be
\eqn\fobcs{
\eqalign{
\psi(\sigma_1+1, \sigma_2) &= e^{2\pi i s/N} \psi(\sigma_1, \sigma_2)
\cr
\psi(\sigma_1+1, \sigma_2) &= -e^{2\pi i s/N} \psi(\sigma_1, \sigma_2)
\cr}
\qquad \eqalign{&{\rm (R)} \cr & {\rm (NS)}~. \cr}
}
The energies of the states are then
\eqn\ener{
\eqalign{
E&=\half ({s\over N})^2 - \half ({s\over N}) + {1\over 3} +m_1 + m_2 {s\over N}
\cr
E&=\half ({s\over N})^2 - {1\over 6} +n_1(\half-{s\over N}) +
n_2 {s\over N}+n_3  \cr}
\qquad \eqalign{&{\rm (R)} \cr & {\rm (NS)} \cr}
}
where $m_i$ and $n_i$ are integers. The GSO projections impose additional
restrictions on these integers. It is sufficient to impose the
condition  \levm\ on the lowest energy states of the left and right
sectors, which
correspond to $m_1=m_2=0$ in the R sector and $n_1=1$, $n_2=n_3=0$ in the
NS sector. This leads to the condition\foot{This condition may be relaxed
if we change the GSO projection so that it mixes with the projection
onto $Z_N$ invariant states, as we see later in this section.} that
$s$ is even. We take
\eqn\levmans{
s=2.
}
Other non-zero values lead to equivalent theories.

The necessity of taking $s=2$  (or even) arises because of
the presence of spacetime fermions. In a theory with
spacetime fermions, a $2\pi$ rotation is not trivial,
so one cannot construct a $Z_N$ orbifold with a group action
whose $N$'th power is a $2\pi$ rotation. Rather one needs
a group action
whose $N$'th power is a $4\pi$ rotation. This is accomplished by setting
$s=2$ in \fobcs.

The boundary conditions in the $k$'th twisted sector are then
\eqn\obcs{
\eqalign{
\psi(\sigma_1+1, \sigma_2) &= e^{4\pi i k/N} \psi(\sigma_1, \sigma_2)
\cr
\psi(\sigma_1+1, \sigma_2) &= -e^{4\pi i k/N} \psi(\sigma_1, \sigma_2)
\cr
X(\sigma_1+1, \sigma_2) &= e^{4\pi i k/N} X(\sigma_1, \sigma_2)~. \cr}
\qquad \eqalign{&{\rm (R)} \cr & {\rm (NS)} \cr & \cr}
}
The usual GSO projection gives the following modular
invariant partition function
\eqn\ffpartit{
{\cal Z}_N = {1\over {4N}}
{1\over {| \eta^2 {\bar \eta}^2 {\rm Im}\tau|^3 }}
\sum_{k,l=0}^{N-1} \sum_{\alpha,\beta \atop \gamma,\delta}
\omega_{\alpha,\beta}(k,l) \bar \omega_{\gamma,\delta}(k,l)
 { { \theta^3 \Bigl[ {\alpha \atop
\beta} \Bigr] \theta\Bigl[ {{\alpha+{{2k}\over N}} \atop {\beta+{{2l} \over
N}}}
\Bigr]
\bar\theta^3 \Bigl[ {\gamma \atop
\delta} \Bigr] \bar\theta\Bigl[ {{\gamma+{{2k}\over N}} \atop
{\delta+{{2l} \over N}}} \Bigr]
} \over {\Bigl|\theta\Bigl[ {{ \ha+{{2k} \over N}} \atop
{\ha+ {{2l} \over N}}} \Bigr] \eta^3\Bigr|^2 }}~.
}
The sum over spin structures corresponds to the sum over
$\alpha, \cdots, \delta = 0, \half$. The coefficients
$\omega_{\alpha \beta}(k,l)$
are $\omega_{1/2, 1/2}= \pm e^{-2\pi i k/N}$, $\omega_{1/2,0}= - 1$,
$\omega_{0, 0}=  1$ and $\omega_{0, 1/2}= -e^{-2\pi i k/N}$. The
$e^{-2\pi i k/N}$ factors arise from the action of $(-1)^F$ on the
vacua in the Neveu-Schwarz and Ramond sectors.

For $N$ odd, this partition function has an interpretation as
a string theory living on a cone with deficit angle $2\pi (1-1/N)$.
To see this we use the simple transformation properties of
\ffpartit under shifts of $2k$ and $2l$ by $N$, and
reorganize the sums over twisted
sectors replacing $2k$ and $2l$ by $k$ and $l$. This yields a
partition function whose bosonic contribution is exactly the
same as in the bosonic partition function \bpartit\ :
\eqn\fopartit{
{\cal Z}_N = {1\over {4N}}
{1\over {| \eta^2 {\bar \eta}^2 {\rm Im}\tau|^3 }}
\sum_{k,l=0}^{N-1} \sum_{\alpha,\beta \atop \gamma,\delta}
\omega'_{\alpha,\beta}(k,l) \bar \omega'_{\gamma,\delta}(k,l)
 { { \theta^3 \Bigl[ {\alpha \atop
\beta} \Bigr] \theta\Bigl[ {{\alpha+{{k}\over N}} \atop {\beta+{{l} \over N}}}
\Bigr]
\bar\theta^3 \Bigl[ {\gamma \atop
\delta} \Bigr] \bar\theta\Bigl[ {{\gamma+{{k}\over N}} \atop
{\delta+{{l} \over N}}} \Bigr]
} \over {\Bigl|\theta\Bigl[ {{ \ha+{{k} \over N}} \atop
{ \ha+{{l} \over N}}} \Bigr] \eta^3\Bigr|^2 }}~.
}
Here we define $\omega'_{\alpha \beta} (k,l) =
\omega_{\alpha \beta}(k/2, l) e^{2\pi i(\alpha l- \beta k)}$.
Using the Riemann theta identity \mumford\ it may be shown that
\fopartit\ agrees
with that calculated in \dabhol\ (equation (2.20)).

To obtain a theory with a thermodynamic interpretation,
spacetime fermions should be periodic around the tip of the
cone. This is analogous to demanding that spacetime fermions be antiperiodic
around the cylinder, as in the usual flat space finite temperature
calculations.
In that case, care must be taken in defining the GSO projection
to ensure that fermions have the correct properties \atick.
It turns out that the fermions described by the
above orbifold are indeed periodic around the cone, so no
additional signs need to be introduced into \fopartit\ to obtain
a thermodynamic partition function.

This model contains fermionic and bosonic states in the left and
right-moving sectors. In the twisted sectors, the projection
onto $Z_N$ invariant states eliminates all fermions. In the untwisted
sector, fermionic states with zero momentum in
the $X_1$ and $X_2$ directions are eliminated by the
$Z_N$ projection, but non-zero momentum states survive.
In each twisted sector, a tachyon is present.
When $k$ is odd, these correspond to the ground state of the
(NS,NS) sector with $M^2=4(-1+k/N)$, which is not projected out in this case.
When $k$ is even, the NS ground state is projected out
and the states
\eqn\tacht{
\psi_{-1/2+k/N} \bar {\tilde  \psi}_{-1/2+k/N} | 0;0\rangle
}
are tachyonic with $M^2= -4 k/N$.
No tachyons are present in the untwisted sector. For $N=1$, this
partition function reproduces that of the usual flat space type II string.

For $N=2N'$, with $N'$ odd or even, \fpartit\ similarly has an interpretation
as a string theory on a cone -- this time with  deficit angle
$2\pi (1-1/N')$. However, it turns out these partition functions
are identical to the theories discussed in the previous subsection, and
all spacetime fermions are projected out.

Finally, the free energy at genus one is given by
\eqn\storamp{
\beta F_N = -V_T \int_{{\cal F}} { {d^2 \tau} \over {({\rm Im} \tau)^2} }
{\cal Z}_N~,
}
where ${\cal Z}_N$ is one of the type II string partition functions discussed
above.
As for the bosonic string, the free energy will be negative infinity
due to the infrared divergence from the tachyons present in the
twisted sectors. This should be dealt with by giving these fields
expectation values, leading to a genus-zero contribution to the free
energy.

Note that in the theory with $N$ odd, \fopartit, the leading
infrared divergence arises from the sectors with $k=1$ and $k=N-1$,
which contain tachyonic states with $M^2=4(-1+1/N)$. It is
natural to presume that
this leading divergence is the same for superstrings
propagating on cones of arbitrary deficit angle. A
consistency check on this is
that in the limit $N\to 1$ one regains the usual
supersymmetric flat space theory. Hence we assume that
analogs of these
states are present for arbitrary (non-integral) $N$, and that
the mass formula may therefore be analytically continued in $N$.
As $N\to 1$, these states become massless, consistent with the
fact that the $N=1$ orbifold is just the usual type II superstring.

In the limit $N\to 1$, it is possible to use effective
field theory methods as in \atick\ to make the arguments concerning
the formation of the condensate more precise. The
effective potential for the two most tachyonic twisted sector fields,
denoted $\phi$ and
$\phi^*$ is
\eqn\effecpot{
V = 4({1\over N} -1) \phi \phi^* +
gu(\phi \phi^*)^2+\cdots
}
which should be valid when the fields are sufficiently small.
Here $g$ is the closed string coupling constant and $u$ is
a constant.
This effective potential may then be minimized for $N>1$
by having $\phi$
jump to non-zero expectation values of order $(N-1)/g$, leading to a
phase transition and a genus-zero contribution to the
free energy.  Unfortunately the non-trivial minimum can not
be found in perturbation theory, and a more detailed description of this
phase transition has not been obtained.

The entropy is related to the free energy by
\eqn\entrop{
S= \beta^2 {\partial F \over \partial \beta}=
-2\pi {\partial F \over \partial N}~.
}
Since, in general, one expects $\vev{\phi} \to 0$ as $N$ approaches 1,
the genus-zero condensate contribution to the free energy should approach
zero faster than $(N-1)$ as indicated by \effecpot. It follows
from \entrop\ that the genus-zero
condensate contribution to the entropy at $N=1$ will vanish.
Thus the details of the physics of this strong coupling
condensate are irrelevant to the calculation of the
genus-zero entropy.

There is an important caveat to the foregoing discussion.
We have described a second order phase transition at
$\epsilon=0$. However corrections of various kinds
could easily perturb this to a first order phase transition, which
might bear on the entropy calculation. Indeed in \atick\ it was
suggested that couplings to the string dilaton alter the
naively second order phase transition at
the Hagedorn temperature to a first order transition at a {\it lower}
critical temperature. A naive translation of this suggestion to
the present context leads to the nonsensical conclusion that a
condensate forms at deficit angles greater than some
critical value  $\epsilon_{crit}<0$, and is therefore
present even in flat space! While this seems unlikely, a better
understanding of the phase transition could certainly alter our
conclusion that the condensate does not enter in to the genus-zero
entropy calculation.


\newsec{Heterotic string}

Now we consider $Z_N$ orbifolds of heterotic string theories
with single conical singularities. In this case no non-trivial
models exist for arbitrary $N$ unless the conical singularity is
accompanied by a Wilson line breaking the symmetry of the internal
gauge group, or some additional twist of the spacetime modes.

We will work with the $E_8 \times E_8$ version of the
heterotic string with a fermionic representation of the gauge
degrees of freedom. In superconformal gauge, the worldsheet action
is then
\eqn\hetact{
S= -{1\over {2\pi }} \int d^2\sigma \bigl( (\partial_\alpha X^\mu)^2-
i \psi^{\mu} \rho^- \partial_- \psi_\mu -
i \lambda^i \rho^+ \partial_+ \lambda^i
- i {\tilde \lambda}^i \rho^+ \partial_+ {\tilde \lambda}^i \bigr)~,
}
where $\mu=0, \cdots, 9$, and $i=1, \cdots,16$. The right-moving
sector is that of the usual RNS string. The left-moving sector
consists of the ten bosonic degrees of freedom plus the two
groups of sixteen fermions which generate the $E_8 \times E_8$ current
algebra. To remove the tachyon, the usual GSO projection is
performed for the right-moving RNS fermions. To correctly generate
the $E_8 \times E_8$ current algebra a GSO--like projection
must be performed on the left-moving fermions, which removes
states with an odd number of $\lambda^i$ and an odd number of
${\tilde \lambda}^i$ oscillators. The states then divide up into
explicit representations of $SO(16)\times SO(16)$.

We are interested in orbifolding this theory by $Z_N$. If we
work in light-cone gauge, the symmetry group of the theory
contains $K=O(8)\times O(16)\times O(16)$, and an element
of $K$ acts on the worldsheet fields as
\eqn\gact{
h \cdot (X,\psi, \lambda, \tilde \lambda) = (\theta X, \theta \psi, \theta_1
\lambda, \theta_2 \tilde \lambda)~.
}
In fact, at the genus-one, we should take an eight-fold cover of $K$
corresponding to the action of $h$ on the different spin structures
of the fermions. If $h$ lies in a $Z_N$ subgroup of $K$, we may write
\eqn\gzn{
\eqalign{
\theta &= (e^{2\pi i r_1/N}, \cdots, e^{2\pi i r_4/N} , \cdots
{\rm c. c.}) \cr
\theta_1 &= (e^{2\pi i p_1/N}, \cdots, e^{2\pi i p_8/N} , \cdots
{\rm c. c.}) \cr
\theta_2 &= (e^{2\pi i q_1/N}, \cdots, e^{2\pi i q_8/N} , \cdots
{\rm c. c.}) \cr}
}
where the $r_i,p_i$ and $q_i$ are integers. As shown by Vafa
\ref\vafa{C. Vafa, \NP {\bf B273} (1986) 592.}, a necessary
condition for modular invariance of the resulting orbifold
is that
\eqn\oddcon{
\sum_1^4 (r_i)^2 = \sum_1^8 (p_i)^2 + (q_i)^2 \quad {\rm mod~N}~,
}
when $N$ is odd, and
\eqn\evencon{
\eqalign{
\sum_1^4 (r_i)^2 &= \sum_1^8 (p_i)^2 + (q_i)^2 \quad {\rm mod~ 2N}\cr
\sum_1^4 r_i &= \sum_1^8 p_i = \sum_1^8 q_i = 0~ {\rm mod~ 2}~, \cr}
}
when $N$ is even.

It is easy to see that if we wish to construct
orbifolds with deficit angles $2\pi (1-1/N)$, we cannot satisfy the
above conditions unless $h$ also has non-trivial action on either
the $\lambda^i, {\tilde \lambda}^i$, or the transverse $X^i$ and $\psi^i$.
Here we consider the case $r_1 = p_1 =1$, with all other $r,p$ and $q$ zero,
which satisfies the above conditions when $N$ is odd.
This corresponds to embedding the spin connection in the first $E_8$ gauge
group, and will lead to an orbifold with an interpretation as a
cosmic string with a Wilson line attached.

The partition function of this theory is
\eqn\hpartit{
\eqalign{
{\cal Z}_N=&{1\over {8N}}
{1\over {| \eta^2 {\bar \eta}^2 {\rm Im}\tau|^3 }}
\sum_{k,l=0}^{N-1}
\sum_{ {{\alpha,\beta\atop \gamma,\delta} \atop \kappa,\lambda}}
\omega'_{\alpha \beta}(k,l)
\bar \rho_{\gamma,\delta}(k,l) \bar \rho_{\kappa, \lambda}(0,0) \cr
& \quad \times
{ {\theta^3 \Bigl[ {\alpha \atop \beta} \Bigr]
 \theta\Bigl[ {{\alpha+{{k}\over N}} \atop {\beta+{{l} \over N}}}
\Bigr] \bar \theta^7 \Bigl[ {\gamma \atop \delta} \Bigr]
\bar \theta \Bigl[ {{\gamma+{{k}\over N}} \atop {\delta+{{l} \over N}}}
\Bigr] \bar \theta^8 \Bigl[ {\kappa \atop \lambda} \Bigr] } \over
{ \Bigl|\theta \Bigl[ {{\ha+{{k}\over N}} \atop {\ha+{{l} \over N}}}
\Bigr]\Bigr|^2 \eta^3 {\bar \eta}^{15} }}~,\cr}
}
where $\omega'(k,l)$ is as defined below \fopartit, and $\bar \rho(k,l)$ are
$\bar \rho_{1/2, 1/2} = \pm e^{i\pi k/N} (-1)^{k+l},$
$\bar \rho_{1/2, 0} =  (-1)^{l}$,
$\bar \rho_{0,0} =1,$ and
$\bar \rho_{0,1/2} = e^{i\pi k/N} (-1)^k$. The sums over
$\alpha, \cdots, \lambda=0,1/2$ correspond to the sum over spin
structures. The genus-one free energy is then
\eqn\hfree{
\beta F =  -V_T \int_{{\cal F}} { {d^2 \tau} \over {({\rm Im} \tau)^2} }
{\cal Z}_N~.
}

This theory contains tachyons in each of the twisted sectors,
localized near the tip of the cone.
When $k$ is odd, the ground state of the NS
sector is not projected out, and the GSO-like projection on the
left-moving fermions removes states with an even number
of the $\lambda^i$ or $\tilde \lambda^i$. The physical states
\eqn\tacho{
\lambda^i_{-1/2} |0;0\rangle~, \quad \tilde \lambda^i_{-1/2} |0;0\rangle~,
\quad \bar \lambda^i_{-1/2} |0;0\rangle~, \quad
\bar {\tilde \lambda}^i_{-1/2} |0;0\rangle~,
}
($i=2,\cdots,8$) are tachyonic with $M^2=4(-1+k/N)$.
When $k$ is even, the tachyonic physical state is
\eqn\tache{
\psi_{-1/2+k/N}  \bar {\tilde \alpha}_{-1+k/N} | 0;0\rangle~,
}
with $M^2= -4 k/N$.
The untwisted sector
contains the usual spectrum of the $E_8 \times E_8$ heterotic
string in flat space, projected onto $Z_N$ invariant states.
The qualitative conclusion that a condensate of these states will
form, leading to a genus-zero contribution to the free energy,
is the same as in the bosonic and type II superstring cases.

\bigskip
{\bf Acknowledgements}

We would like to thank L. Thorlacius for useful discussions.
This work was supported in part by NSF grant PHY-91-16964 and
DOE grant 91-ER40618.

\listrefs
\end